\documentclass[10pt,twocolumn,superscriptaddress,amssymb,amsmath,aps,prl]{revtex4-2}
\usepackage{graphicx}

\usepackage{bbm}

\usepackage[normalem]{ulem}

\usepackage{color}

\begin{document}


\title{Perihelion precession of planetary orbits solved from quantum field theory}
\date{June 17, 2025}
\author{Mikko Partanen}
\affiliation{Photonics Group, Department of Electronics and Nanoengineering, 
Aalto University, P.O. Box 13500, 00076 Aalto, Finland}
\author{Jukka Tulkki}
\affiliation{Engineered Nanosystems Group, School of Science, Aalto University, 
P.O. Box 12200, 00076 Aalto, Finland}

\begin{abstract}
We derive the perihelion precession of planetary orbits using quantum field theory extending the Standard Model to include gravity. Modeling the gravitational bound state of an electron via the Dirac equation of unified gravity [Rep.~Prog.~Phys.~\textbf{88}, 057802 (2025)], and taking the classical planetary state limit, we obtain orbital dynamics exhibiting a precession in agreement with general relativity. This demonstrates that key general relativistic effects in planetary motion can emerge directly from quantum field theory without invoking the geometric framework of general relativity.
\end{abstract}

\maketitle


\section{Introduction}
\vspace{-0.3cm}

\begin{figure}
\centering
\includegraphics[width=0.98\columnwidth]{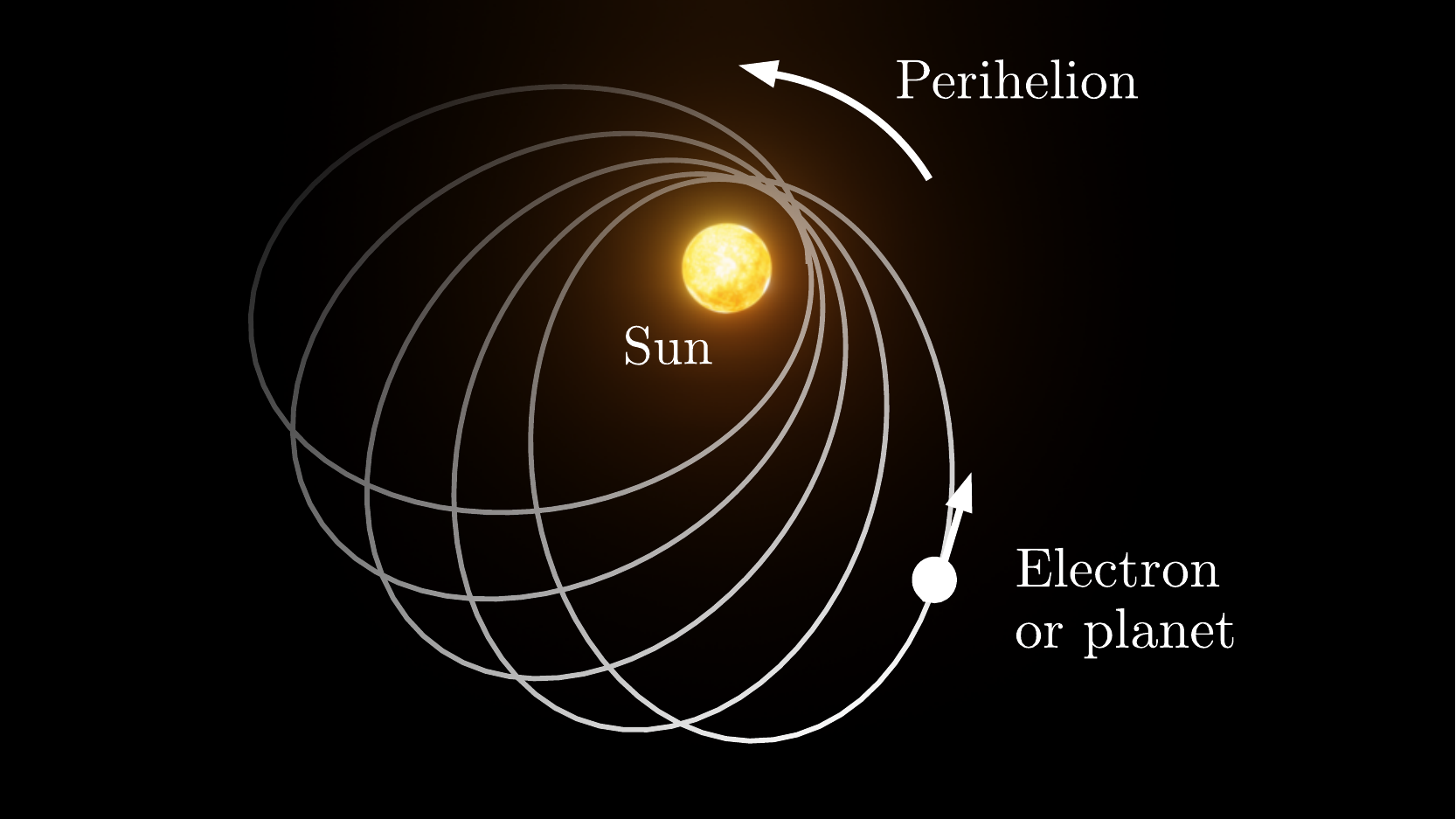}
\caption{\label{fig:illustration}
Illustration of the perihelion precession of an object orbiting the sun. In the present work, we use the Dirac equation of the electron in the presence of the gravity gauge field to calculate the precession of the orbit. The results generalize to larger masses like planets.}
\end{figure}

The perihelion precession of planetary orbits \cite{Will2018,Misner1973,Moore2013}, illustrated in Fig.~\ref{fig:illustration} and most famously observed in the orbit of Mercury \cite{Verrier1859}, stands as one of the classical tests of general relativity (GR) \cite{Einstein1915,Einstein1916,Will1993}. Einstein's theory accurately predicts the observed anomaly in Mercury’s precession, which could not be accounted for by Newtonian mechanics alone. GR provides a geometric description of gravity as the curvature of spacetime, depicted by the metric. There are also equivalent formulations, modifications, and different theories based on the metric, torsion, or nonmetricity \cite{Bahamonde2023a,Aldrovandi2012,Maluf2013,Bhattacharya2017,Damour1996,Brans1961,Chiba2007,Farrugia2016,Sultana2012}, which are all geometric concepts. Alternative approaches based on different ideas, such as string theory \cite{Becker2007,Dine2007,Green1987}, loop quantum gravity \cite{Ashtekar1986,Jacobson1988,Rovelli1990,Rovelli2008}, asymptotic safety \cite{Addazi2022,Niedermaier2006}, noncommutative geometry \cite{Connes1994}, and causal dynamical triangulation \cite{Loll2020} are being developed. However, their ability to provide unique, testable predictions about gravity is typically limited and under active development \cite{Addazi2022}. Therefore, a complete unification of the gravitational interaction with quantum theory has remained one of the outstanding challenges in fundamental physics. Whether gravitational interactions needs to be quantized has also been discussed until recently \cite{Oppenheim2023a,Oppenheim2023b}.

In this work, we derive the perihelion precession of planetary orbits by analyzing localized bound states of electrons governed by the Dirac equation of unified gravity (UG) \cite{Partanen2025a}. The gauge theory of UG is an extension of the Standard Model of particle physics and it incorporates gravitational interaction through a gravity gauge field derived from four one-dimensional unitary gauge symmetries. Therefore, rather than relying on geometric concepts such as spacetime curvature, torsion, or nonmetricity, UG treats gravity as a fundamental interaction mediated by a field, consistent with the structure of relativistic quantum mechanics \cite{Schwartz2014,Peskin2018}. Two alternative formulations of UG have been presented based on different geometric conditions \cite{Partanen2025a}. One condition leads to teleparallel equivalent of GR \cite{Bahamonde2023a,Aldrovandi2012,Maluf2013,Partanen2025a}, and the other condition relies on the Minkowski metric and preserves the original gauge symmetries of the theory. In this work, UG consistently refers to the latter formulation.

Gravitational bound states of electrons are studied by using the Dirac equation of UG in the presence of the static gravity gauge field solved for a point mass. Taking the classical localized planetary state limit of the Dirac equation, we obtain a precession of the orbital motion, which matches the perihelion shift observed in planetary dynamics. This approach demonstrates that the key prediction traditionally attributed to GR can be recovered directly from a quantum field theory formulation of matter interacting via a gravitational field. For the deflection of light in a gravitational field, a corresponding study has been presented in a separate preprint \cite{Partanen2025c}. Detailed comparison of the difference between UG and GR is left as a topic of further work.

\section{Gravity gauge field}
\vspace{-0.3cm}

We follow a semiclassical approach, in which the gravitational field is treated classically. Thus, our approach reminds the conventional calculation of the bound states of electrons in the classical electromagnetic potential of the atomic nucleus in quantum electrodynamics (QED) \cite{Feynman1961,Landau1982}. We start our description with a brief review of the solution of the UG gravity gauge field for a classical point mass as presented in Ref.~\cite{Partanen2025c}.

We use the global Cartesian Minkowski frame coordinates $x^\mu=(ct,x,y,z)$, where $c$ is the speed of light in vacuum and at zero gravitational potential. For the diagonal Minkowski metric tensor $\eta_{\mu\nu}$, we use the sign convention $\eta_{00}=1$ and $\eta_{xx}=\eta_{yy}=\eta_{zz}=-1$. The Einstein summation convention is used for all repeated indices in this work. In the global Minkowski frame, the gravitational field equation of UG in the harmonic gauge is given by \cite{Partanen2025a}
\begin{equation}
 -P^{\mu\nu,\rho\sigma}\partial^2H_{\rho\sigma}=\kappa T_\mathrm{m}^{\mu\nu}.
 \label{eq:UGgravity}
\end{equation}
Here $H_{\rho\sigma}$ is the gravity gauge field, $\partial^2=\partial^\rho\partial_\rho$ is the d'Alembert operator, and $\kappa=8\pi G/c^4$ is Einstein's constant, in which $G$ is the gravitational constant, and $P^{\mu\nu,\rho\sigma}=\frac{1}{2}(\eta^{\mu\sigma}\eta^{\rho\nu}+\eta^{\mu\rho}\eta^{\nu\sigma}-\eta^{\mu\nu}\eta^{\rho\sigma})$. The source term of gravity, $T_\mathrm{m}^{\mu\nu}$, on the right in Eq.~\eqref{eq:UGgravity}, is the stress-energy-momentum tensor of matter and vector gauge fields, such as the electromagnetic field. The harmonic gauge is given by $P^{\mu\nu,\rho\sigma}\partial_\rho H_{\mu\nu}=0$.

We consider the solution of Eq.~\eqref{eq:UGgravity} for the stress-energy-momentum tensor of a mass $M$ located at the origin, given by $T_\mathrm{m}^{\mu\nu}=Mc^2\delta(\mathbf{r})\delta_0^\mu\delta_0^\nu$. Here $\delta(\mathbf{r})$ is the three-dimensional Dirac delta function, $\mathbf{r}=(x,y,z)$, and $\delta_\nu^\mu$ is the Kronecker delta. The mass $M$ is assumed to be large, like the mass of the sun, so that the mass of the electron, studied below, is negligible in the calculation of the total gravitational field. Using $T_\mathrm{m}^{\mu\nu}$, the exact solution of Eq.~\eqref{eq:UGgravity} is given by \cite{Partanen2025a}
\begin{equation}
 H_{\mu\nu}=\left[
 \begin{array}{cccc}
 \frac{\Phi}{c^2} & 0 & 0 & 0\\
 0 & \frac{\Phi}{c^2} & 0 & 0\\
 0 & 0 & \frac{\Phi}{c^2} & 0\\
 0 & 0 & 0 & \frac{\Phi}{c^2}
 \end{array}\right],\hspace{0.5cm}\Phi=-\frac{GM}{r}.
 \label{eq:H}
\end{equation}
Here $\Phi$ is the Newtonian gravitational potential satisfying $\nabla^2\Phi=4\pi GM\delta(\mathbf{r})$, where $\nabla=(\partial_x,\partial_y,\partial_z)$ is the three-dimensional gradient operator, and $r=|\mathbf{r}|=\sqrt{x^2+y^2+z^2}$. The constants of integration have been set to zero by assuming that the gravitational field vanishes at infinity.

\section{Dirac equation under gravity}
\vspace{-0.3cm}

In UG, all dynamical equations are formulated directly within the global Minkowski spacetime, and the gravitational interaction enters explicitly through a gauge field \cite{Partanen2025a}. This contrasts fundamentally with GR, where gravitational effects are encoded implicitly via the spacetime metric and the geometry it defines \cite{Misner1973}. This raises the question whether UG is capable of describing all observable phenomena of gravitational interaction, which are interpreted to support GR. Next, we study the gravitational bound state of an electron via the Dirac equation of UG \cite{Partanen2025a}, approximately diagonalize the Hamiltonian system, and then consider its classical planetary orbit limit.

The Dirac equation of UG has been derived from the Lagrangian density through the well-known Euler--Lagrange equations \cite{Partanen2025a}. In the presence of the electromagnetic and gravitational fields, it is given by
\begin{align}
 &i\hbar c\boldsymbol{\gamma}^\rho\vec{\partial}_\rho\psi\!-\!m_\mathrm{e}c^2\psi
 \!=q_\mathrm{e}c\boldsymbol{\gamma}^\rho\psi A_\rho\nonumber\\
 &+P^{\mu\nu,\rho\sigma}\Big(i\hbar c\boldsymbol{\gamma}_\sigma\vec{\partial}_\rho\psi-\frac{m_\mathrm{e}c^2}{2}\eta_{\rho\sigma}\psi+\frac{i\hbar c}{2}\boldsymbol{\gamma}_\sigma\psi\vec{\partial}_\rho\nonumber\\
 &-q_\mathrm{e}c\boldsymbol{\gamma}_\sigma\psi A_\rho+\frac{q_\mathrm{e}c}{4}\eta_{\rho\sigma}\boldsymbol{\gamma}^\lambda A_\lambda\psi\Big)H_{\mu\nu}.
 \label{eq:DiracUGM}
\end{align}
Here $\hbar$ is the reduced Planck constant, $m_\mathrm{e}$ is the electron mass, $q_\mathrm{e}$ is the electron charge, $\psi$ is the Dirac field describing the electron, $A_\mu$ is the electromagnetic four-potential, and $\boldsymbol{\gamma}^\mu$ are the conventional $4\times 4$ Dirac gamma matrices.

In the present work, we study the electron at zero electromagnetic four-potential and accordingly set $A_\mu=0$. Substituting the gravity gauge field $H_{\mu\nu}$ from  Eq.~\eqref{eq:H} into Eq.~\eqref{eq:DiracUGM}, and accounting for the time independence of $H_{\mu\nu}$ by setting its time derivatives to zero, we obtain
\begin{align}
 \Big(1-\frac{\Phi}{c^2}\Big)m_\mathrm{e}c^2\boldsymbol{\beta}\psi
 +c\boldsymbol{\alpha}\cdot\hat{\mathbf{p}}\psi
 =i\hbar\Big(1-\frac{2\Phi}{c^2}\Big)\frac{\partial \psi}{\partial t}.
 \label{eq:DiracUGM4}
\end{align}
Here we have used the well-known alpha and beta matrices of the Dirac theory, given by $\boldsymbol{\beta}=\boldsymbol{\gamma}^0$ and $\boldsymbol{\alpha}^i=\boldsymbol{\gamma}^0\boldsymbol{\gamma}^i$, $i\in\{x,y,z\}$, and the three-component vector of the alpha matrices, given by $\boldsymbol{\alpha}=(\boldsymbol{\alpha}^x,\boldsymbol{\alpha}^y,\boldsymbol{\alpha}^z)$. We have also used the definition of the three-dimensional momentum operator, given by $\hat{\mathbf{p}}=-i\hbar\nabla$.

Dividing Eq.~\eqref{eq:DiracUGM4} by $1-2\Phi/c^2$, we can then rewrite the Dirac equation of UG in the Hamiltonian form, given by
\begin{equation}
 \begin{array}{c}
 \hat{H}\psi=i\hbar\dfrac{\partial \psi}{\partial t},
 \hspace{0.5cm}\hat{H}=C_1m_\mathrm{e}c^2\boldsymbol{\beta}+C_2c\boldsymbol{\alpha}\cdot\hat{\mathbf{p}},\\
 C_1=\dfrac{1-\frac{\Phi}{c^2}}{1-\frac{2\Phi}{c^2}},\hspace{0.5cm}C_2=\dfrac{1}{1-\frac{2\Phi}{c^2}}.
 \label{eq:Hamiltonian}
 \end{array}
\end{equation}
Here $\hat{H}$ is the Hamiltonian operator and we have defined the coefficients $C_1$ and $C_2$ to contain the dependencies on the gravitational potential. The conventional Dirac Hamiltonian in free space is recovered at zero gravitational potential, for which the values of $C_1$ and $C_2$ are equal to unity. In the general case, the non-unity  values of these coefficients have fundamental consequences, such as the precession of orbits encountered in the classical planetary state limit as derived below.

\section{\label{sec:diagonalization}Approximate diagonalization}
\vspace{-0.3cm}

We could solve the Dirac equation of UG in Eqs.~\eqref{eq:DiracUGM}--\eqref{eq:Hamiltonian} numerically for a given initial wavepacket state, but, to obtain a deeper insight into the classical planetary state limit, we look below for an analytic approximate solution. We approximately diagonalize the Hamiltonian operator in Eq.~\eqref{eq:Hamiltonian} by utilizing the Foldy--Wouthuysen transformation \cite{Foldy1950,Foldy1952,Landau1982}. The Foldy--Wouthuysen transformation is a unitary transformation on the Dirac spinor, $\psi\rightarrow \psi'=U\psi$, given by
\begin{equation}
 U=e^{\theta\boldsymbol{\beta}\boldsymbol{\alpha}\cdot\hat{\mathbf{p}}/|\mathbf{p}|}=\cos\theta\,\mathbf{I}_4+\frac{\sin\theta}{|\mathbf{p}|}\boldsymbol{\beta}\boldsymbol{\alpha}\cdot\hat{\mathbf{p}}.
 \label{eq:U}
\end{equation}
Here $\mathbf{I}_4$ is a $4\times4$ identity matrix and $\theta$ is the transformation angle, which can be chosen arbitrarily. Our only approximation in the application of the Foldy--Wouthuysen transformation below is that we neglect the commutator of the momentum operator and the Newtonian potential by setting $[\hat{\mathbf{p}},C_1]=-i\hbar(\partial C_1/\partial\Phi)\nabla\Phi\rightarrow 0$ and $[\hat{\mathbf{p}},C_2]=-i\hbar(\partial C_2/\partial\Phi)\nabla\Phi\rightarrow 0$. This is well justified since our goal is to derive the classical limit of the quantum theory \cite{Shankar1994,Sakurai1994}. Alternatively, we could explicitly keep these commutators in our equations and set the Planck constant to zero when taking the classical planetary state limit at the end. This procedure naturally eliminates all quantum corrections.

The Hamiltonian operator is transformed in the Foldy--Wouthuysen transformation as $\hat{H}\rightarrow\hat{H}'=U\hat{H}U^{-1}$. After some algebra using Eqs.~\eqref{eq:Hamiltonian} and \eqref{eq:U} and the commutativity properties of the Dirac matrices, we obtain
\begin{align}
 \hat{H}' &=\Big[C_2c\cos(2\theta)-\frac{C_1m_\mathrm{e}c^2}{|\mathbf{p}|}\sin(2\theta)\Big]\boldsymbol{\alpha}\cdot\hat{\mathbf{p}}\nonumber\\
 &\hspace{0.4cm}+\Big[C_1m_\mathrm{e}c^2\cos(2\theta)+C_2c|\mathbf{p}|\sin(2\theta)\Big]\boldsymbol{\beta}.
 \label{eq:Hprime}
\end{align}
Note that the equality holds when the commutators $[\hat{\mathbf{p}},C_1]$ and $[\hat{\mathbf{p}},C_2]$ are set to zero according to the discussion above. Within the same approximation, we choose the transformation angle $\theta$ so that the following relations are satisfied:
\begin{align}
 \sin(2\theta) &=\frac{C_2|\mathbf{p}|}{\sqrt{C_1^2m_\mathrm{e}^2c^2+C_2^2\mathbf{p}^2}},\nonumber\\
 \cos(2\theta) &=\frac{C_1m_\mathrm{e}c}{\sqrt{C_1^2m_\mathrm{e}^2c^2+C_2^2\mathbf{p}^2}}.
 \label{eq:angles}
\end{align}
After substituting $\sin(2\theta)$ and $\cos(2\theta)$ from Eq.~\eqref{eq:angles} into Eq.~\eqref{eq:Hprime}, we straightforwardly obtain
\begin{equation}
 \hat{H}'=c\sqrt{C_1^2m_\mathrm{e}^2c^2+C_2^2\mathbf{p}^2}\boldsymbol{\beta}.
\end{equation}
Therefore, through the Foldy--Wouthuysen transformation in Eq.~\eqref{eq:U}, the Dirac equation in Eq.~\eqref{eq:Hamiltonian} becomes $\hat{H}'\psi'=i\hbar\partial\psi'/\partial t$.
Since, $\hat{H}'$ is independent of time, we furthermore obtain
\begin{equation}
 \hat{H}^{\prime 2}\psi'=(C_1^2m_\mathrm{e}^2c^4+C_2^2c^2\mathbf{p}^2)\psi'=\hat{E}^2\psi'.
 \label{eq:H2}
\end{equation}
Here $\hat{E}=i\hbar\partial/\partial t$ is the energy operator. Equation \eqref{eq:H2} contains only scalar quantities in front of the Dirac spinors, and we can use it to obtain the classical planetary state limit as discussed below.

\section{\label{seq:orbit}Planetary orbit dynamics}
\vspace{-0.3cm}

Next, we take the classical limit of Eq.~\eqref{eq:H2} through the expectation value. We calculate the expectation value of the squared Hamiltonian $\hat{H}^{\prime 2}$ by multiplying Eq.~\eqref{eq:H2} by $\psi^{\prime\dag}$ from the left and by integrating over the volume. Thus, writing $C_1=C_1(r)$ and $C_2=C_2(r)$, we obtain
\begin{equation}
 \langle[C_1(r)]^2\rangle m_\mathrm{e}^2c^4+c^2\langle[C_2(r)]^2\hat{\mathbf{p}}^2\rangle=\langle\hat{E}^2\rangle.
 \label{eq:expectation}
\end{equation}
Above, the Dirac spinor $\psi'$ has not been specified. Next, we assume that the electron is in an initial state that resembles states of classical mechanics, i.e., states with well-defined position and momentum \cite{Shankar1994}. Simultaneous eigenstates of position and momentum do not exist, but an approximate eigenstate can be written as \cite{Shankar1994}
\begin{equation}
 \psi'=\Big(\frac{1}{\pi(\Delta\mathbf{r})^2}\Big)^{3/4}e^{i\langle\hat{\mathbf{p}}\rangle\cdot\mathbf{r}/\hbar}e^{-(\mathbf{r}-\langle\mathbf{r}\rangle)^2/[2(\Delta\mathbf{r})^2]}.
 \label{eq:eigenstate}
\end{equation}
Here $\langle\mathbf{r}\rangle$ and $\langle\hat{\mathbf{p}}\rangle$ are the position and momentum expectation values, respectively, and $\Delta\mathbf{r}$ is the standard deviation of position. The Standard deviation of momentum is $\Delta\hat{\mathbf{p}}=\hbar/(\sqrt{2}\Delta\mathbf{r})$. The standard deviations of position and momentum can both be taken negligibly small in the macroscopic scale \cite{Shankar1994}. This certainly applies in the astrophysical scale of current interest. For the approximate eigenstate in Eq.~\eqref{eq:eigenstate}, we can write $\langle[C_1(r)]^2\rangle=[C_1(\langle r\rangle)]^2$, $\langle[C_2(r)]^2\hat{\mathbf{p}}^2\rangle=[C_2(\langle r\rangle)]^2\langle\hat{\mathbf{p}}\rangle^2$, and $\langle\hat{\mathbf{p}}\rangle=m_\mathrm{e}\langle\hat{\mathbf{v}}\rangle$, where $\hat{\mathbf{v}}$ is the velocity expectation value. Then, from Eq.~\eqref{eq:expectation}, after a division by $m_\mathrm{e}^2c^2$ and denoting the energy per unit mass by $\mathcal{E}=\langle\hat{E}\rangle/m_\mathrm{e}$, we obtain \emph{the orbit equation} of UG, given by
\begin{equation}
 [C_1(\langle r\rangle)]^2c^2+[C_2(\langle r\rangle)]^2\langle\hat{\mathbf{v}}\rangle^2=\frac{\mathcal{E}^2}{c^2}.
 \label{eq:expectation2}
\end{equation}

Next, we express the squared velocity expectation value $\langle\hat{\mathbf{v}}\rangle^2$ in polar coordinates $(r,\phi)$ for motion in the $xy$ plane by writing $\langle\hat{\mathbf{v}}\rangle^2=(dr/dt)^2+(rd\phi/dt)^2$. For brevity, we omit the angle brackets by writing $\langle r\rangle=r$ and $\langle\phi\rangle=\phi$. Also, using the orbital angular momentum per unit mass, denoted by $\mathcal{J}=-r^2d\phi/dt$, which is a constant of motion for the isolated system, from the orbit equation in Eq.~\eqref{eq:expectation2}, we obtain
\begin{equation}
 [C_1(r)]^2c^2+[C_2(r)]^2\Big[\Big(\frac{dr}{dt}\Big)^2+\frac{\mathcal{J}^2}{r^2}\Big]=\frac{\mathcal{E}^2}{c^2}.
 \label{eq:expectation3}
\end{equation}

Next, we simplify the orbit equation in Eq.~\eqref{eq:expectation3} by performing the change of variables, given by
\begin{equation}
 r(\phi)=\frac{1}{u(\phi)}-\frac{GM}{c^2}.
 \label{eq:rvariable}
\end{equation}
The radial velocity, $dr/dt$, in Eq.~\eqref{eq:expectation3}, can be rewritten using the chain rule as
\begin{align}
 \frac{dr}{dt} &=\frac{dr}{d\phi}\frac{d\phi}{dt}=-\frac{1}{u^2}\frac{du}{d\phi}\frac{d\phi}{dt}
 =\frac{\mathcal{J}}{u^2r^2}\frac{du}{d\phi}
 =\frac{\mathcal{J}}{\big(1-\frac{GMu}{c^2}\big)^2}\frac{du}{d\phi}.
 \label{eq:substitute1}
\end{align}
The coefficients $[C_1(r)]^2$ and $[C_2(r)]^2$ in Eq.~\eqref{eq:expectation3} are rewritten using Eqs.~\eqref{eq:H}, \eqref{eq:Hamiltonian}, and \eqref{eq:rvariable} as
\begin{align}
 [C_1(r)]^2 &=\frac{\big(1+\frac{GM}{c^2r}\big)^2}{\big(1+\frac{2GM}{c^2r}\big)^2}
 =\frac{1}{\big(1+\frac{GMu}{c^2}\big)^2},\nonumber\\
 [C_2(r)]^2 &=\frac{1}{\big(1+\frac{2GM}{c^2r}\big)^2}
 =\frac{\big(1-\frac{GMu}{c^2}\big)^2}{\big(1+\frac{GMu}{c^2}\big)^2}.
\end{align}
Correspondingly, the term $\mathcal{J}^2/r^2$ in in the second term on the left in Eq.~\eqref{eq:expectation3} is rewritten as
\begin{equation}
 \frac{\mathcal{J}^2}{r^2}=\frac{\mathcal{J}^2u^2}{\big(1-\frac{GMu}{c^2}\big)^2}.
 \label{eq:substitute3}
\end{equation}
Using Eqs.~\eqref{eq:substitute1}--\eqref{eq:substitute3}, the orbit equation of UG in Eq.~\eqref{eq:expectation3} becomes
\begin{equation}
 \frac{1}{\big(1+\frac{GMu}{c^2}\big)^2}\Bigg\{c^2+\mathcal{J}^2\Bigg[\frac{\big(\frac{du}{d\phi}\big)^2}{\big(1-\frac{GMu}{c^2}\big)^2}+u^2\Bigg]\Bigg\}=\frac{\mathcal{E}^2}{c^2}.
 \label{eq:expectation4}
\end{equation}

Taking the derivative of Eq.~\eqref{eq:expectation4} with respect to $\phi$, noting that energy and orbital angular momentum are constants of motion, and thus, independent of $\phi$, multiplying the resulting equation by $\big(1-G^2M^2u^2/c^4\big)^2/[2\mathcal{J}^2du/d\phi]$, and rearranging the terms, we obtain
\begin{align}
 &\frac{d^2u}{d\phi^2}+\frac{\big(1-\frac{GMu}{c^2}\big)^3\!+\!\frac{2G^2M^2}{c^4}\big(\frac{du}{d\phi}\big)^2}{1-\frac{G^2M^2u^2}{c^4}}u
 =\frac{\big(1-\frac{GMu}{c^2}\big)^2}{1+\frac{GMu}{c^2}}\frac{GM}{\mathcal{J}^2}.
 \label{eq:Binet}
\end{align}
Equation \eqref{eq:Binet} cannot be solved analytically in a closed form. Therefore, in the general case, it must be solved numerically. However, perturbative solutions can be derived in the lowest powers of the gravitational constant. Taking the Taylor series of the terms of Eq.~\eqref{eq:Binet} in powers of $GMu/c^2$, dropping out the second- and higher-order terms, and rearranging the terms, Eq.~\eqref{eq:Binet} becomes
\begin{equation}
 \frac{d^2u}{d\phi^2}+u=\frac{GM}{\mathcal{J}^2}+\frac{3GMu^2}{c^2}.
 \label{eq:Binet2}
\end{equation}
Equation \eqref{eq:Binet2} is equivalent in form to the Binet equation, where the terms apart from the last term describe Newtonian orbits, and the last term is known as the GR correction \cite{Misner1973,Moore2013}. Here we have significantly derived Eq.~\eqref{eq:Binet2} without a curved metric or other concepts of GR.

Based on previous literature \cite{Moore2013,Misner1973}, a perturbative solution of Eq.~\eqref{eq:Binet2}, describing an elliptical orbit with precession, is known to be
\begin{equation}
 u(\phi)\approx\frac{1}{a(1-e^2)}\Big\{1+e\cos\Big[\Big(1-\frac{\Delta\phi}{2\pi}\Big)\phi-\phi_0\Big]\Big\}.
 \label{eq:usolution}
 \end{equation}
Here the eccentricity $e$ and the phase angle $\phi_0$ are integration constants, and $a=\mathcal{J}^2/[GM(1-e^2)]$ is the semimajor axis. The perihelion shift $\Delta\phi$ is given by \cite{Misner1973,Moore2013}
\begin{equation}
 \Delta\phi=\frac{6\pi GM}{c^2a(1-e^2)}.
\end{equation}
Substituting the solution of the Binet equation in Eq.~\eqref{eq:usolution} into Eq.~\eqref{eq:rvariable}, we obtain the radial coordinate of the orbit in terms of the polar angle as
\begin{equation}
 r(\phi)\approx\frac{a(1-e^2)}{1+e\cos\big[\big(1-\frac{\Delta\phi}{2\pi}\big)\phi-\phi_0\big]}-\frac{GM}{c^2}.
 \label{eq:ellipse}
\end{equation}
This is the equation for an ellipse with a perihelion precession in agreement with GR. The result of GR exactly agrees with our result when the calculation is performed in the isotropic coordinates of the Schwarzschild metric. Here we have derived Eq.~\eqref{eq:ellipse} assuming the electron described by the Dirac equation of UG. However, since Eq.~\eqref{eq:ellipse} is independent of the electron mass, our results can be generalized for arbitrary masses that are only restricted to be small in comparison with the large mass $M$, whose gravitational field governs the motion of the smaller mass. This suggests that our results generalize for planets orbiting the sun and demonstrates the predictive power of UG, originally based on the gauge symmetry principles of elementary particles \cite{Partanen2025a}. Detailed study of the higher-order differences in orbital dynamics between UG and GR is left as a topic of further work.

\section{Conclusion}
\vspace{-0.3cm}

We have solved the perihelion precession of planetary orbits, one of the key predictions of GR, using quantum field theory as formulated in UG \cite{Partanen2025a}. Our results affirm that UG is a physically predictive extension of the Standard Model, whose classical limit directly recovers classical gravitational phenomena. This study reinforces the consistency between UG and known observations on gravitational interaction. We expect that UG provides a solid platform for exploring higher-order deviations from GR in orbital dynamics both in the classical and quantum physics regimes. Since UG and GR do not contain free parameters, the proposed measurements of second-order gravitational effects in powers of the gravitational constant are expected to enable experimentally distinguishing between these theories in the near future \cite{Turyshev2007,Turyshev2009b,Partanen2025c}. In the lowest order in the gravitational constant studied in this work, the results of UG and GR are equal. In the case of UG, the results follow without using the curved metric or other geometric concepts of GR. This raises the question if gravitation can be reduced to the metric in its most general form.

\section{Acknowledgments}
\vspace{-0.3cm}
\begin{acknowledgments}
This work has been funded by the Research Council of Finland under Contract No.~349971.
\end{acknowledgments}


\begin{thebibliography}{10}
\newcommand{\enquote}[1]{``#1''}

\bibitem{Will2018}
C.~M. Will, \enquote{New general relativistic contribution to {M}ercury's
  perihelion advance,} \emph{Phys. Rev. Lett.} \textbf{120}, 191101 (2018).

\bibitem{Misner1973}
C.~W. Misner, K.~S. Thorne, and J.~A. Wheeler, \emph{Gravitation}, Freeman, New
  York (1973).

\bibitem{Moore2013}
T.~A. Moore, \emph{A General Relativity Workbook}, University Science Books,
  Mill Walley, CA (2013).

\bibitem{Verrier1859}
U.~J. Le~Verrier, \enquote{Theorie du mouvement de {M}ercure,} \emph{Annales de
  l’Observatoire de Paris} \textbf{5}, 1 (1859).

\bibitem{Einstein1915}
A.~Einstein, \emph{Erklärung der {P}erihelbewegung des {M}erkur aus der
  allgemeinen {R}elativitätstheorie}, 831--839 (1915).

\bibitem{Einstein1916}
A.~Einstein, \enquote{Die {G}rundlage der allgemeinen {R}elativitätstheorie,}
  \emph{Ann. Phys.} \textbf{354}, 769 (1916).

\bibitem{Will1993}
C.~M. Will, \emph{Theory and experiment in gravitational physics}, Cambridge
  University Press, Cambridge (1993).

\bibitem{Bahamonde2023a}
S.~Bahamonde, K.~F. Dialektopoulos, C.~Escamilla-Rivera, G.~Farrugia, V.~Gakis,
  M.~Hendry, M.~Hohmann, J.~L. Said, J.~Mifsud, and E.~D. Valentino,
  \enquote{Teleparallel gravity: from theory to cosmology,} \emph{Rep. Prog.
  Phys.} \textbf{86}, 026901 (2023).

\bibitem{Aldrovandi2012}
R.~Aldrovandi and J.~G. Pereira, \emph{Teleparallel Gravity: An Introduction},
  Springer, Dordrecht (2012).

\bibitem{Maluf2013}
J.~W. Maluf, \enquote{The teleparallel equivalent of general relativity,}
  \emph{Ann. Phys.} \textbf{525}, 339 (2013).

\bibitem{Bhattacharya2017}
S.~Bhattacharya and S.~Chakraborty, \enquote{Constraining some {H}orndeski
  gravity theories,} \emph{Phys. Rev. D} \textbf{95}, 044037 (2017).

\bibitem{Damour1996}
T.~Damour and G.~Esposito-Far\`ese, \enquote{Tensor-scalar gravity and
  binary-pulsar experiments,} \emph{Phys. Rev. D} \textbf{54}, 1474 (1996).

\bibitem{Brans1961}
C.~Brans and R.~H. Dicke, \enquote{Mach's principle and a relativistic theory
  of gravitation,} \emph{Phys. Rev.} \textbf{124}, 925 (1961).

\bibitem{Chiba2007}
T.~Chiba, T.~L. Smith, and A.~L. Erickcek, \enquote{Solar system constraints to
  general {$f(R)$} gravity,} \emph{Phys. Rev. D} \textbf{75}, 124014 (2007).

\bibitem{Farrugia2016}
G.~Farrugia, J.~L. Said, and M.~L. Ruggiero, \enquote{Solar system tests in
  {$f(T)$} gravity,} \emph{Phys. Rev. D} \textbf{93}, 104034 (2016).

\bibitem{Sultana2012}
J.~Sultana, D.~Kazanas, and J.~L. Said, \enquote{Conformal {W}eyl gravity and
  perihelion precession,} \emph{Phys. Rev. D} \textbf{86}, 084008 (2012).

\bibitem{Becker2007}
K.~Becker, M.~Becker, and J.~Schwarz, \emph{String theory and {M}-theory: A
  modern introduction}, Cambridge University Press, Cambridge (2007).

\bibitem{Dine2007}
M.~Dine, \emph{Supersymmetry and String Theory}, Cambridge University Press,
  Cambridge (2007).

\bibitem{Green1987}
M.~B. Green, J.~H. Schwarz, and E.~Witten, \emph{Superstring Theory: Volume 1,
  Introduction}, Cambridge University Press, Cambridge (1987).

\bibitem{Ashtekar1986}
A.~Ashtekar, \enquote{New variables for classical and quantum gravity,}
  \emph{Phys. Rev. Lett.} \textbf{57}, 2244 (1986).

\bibitem{Jacobson1988}
T.~Jacobson and L.~Smolin, \enquote{Nonperturbative quantum geometries,}
  \emph{Nucl. Phys. B} \textbf{299}, 295 (1988).

\bibitem{Rovelli1990}
C.~Rovelli and L.~Smolin, \enquote{Loop space representation of quantum general
  relativity,} \emph{Nucl. Phys. B} \textbf{331}, 80 (1990).

\bibitem{Rovelli2008}
C.~Rovelli, \enquote{Loop quantum gravity,} \emph{Living Rev. Relativ.}
  \textbf{11}, 5 (2008).

\bibitem{Addazi2022}
{A. Addazi \emph{et al.}}, \enquote{Quantum gravity phenomenology at the dawn
  of the multi-messenger era—a review,} \emph{Prog. Part. Nucl. Phys.}
  \textbf{125}, 103948 (2022).

\bibitem{Niedermaier2006}
M.~Niedermaier and M.~Reuter, \enquote{The asymptotic safety scenario in
  quantum gravity,} \emph{Living Rev. Relativ.} \textbf{9}, 5 (2006).

\bibitem{Connes1994}
A.~Connes, \emph{Non-commutative geometry}, Academic Press, Boston, MA (1994).

\bibitem{Loll2020}
R.~Loll, \enquote{Quantum gravity from causal dynamical triangulations: a
  review,} \emph{Class. Quantum Grav.} \textbf{37}, 013002 (2019).

\bibitem{Oppenheim2023a}
J.~Oppenheim, \enquote{Is it time to rethink quantum gravity?} \emph{Int. J.
  Mod. Phys. D} \textbf{32}, 2342024 (2023).

\bibitem{Oppenheim2023b}
J.~Oppenheim, \enquote{A postquantum theory of classical gravity?} \emph{Phys.
  Rev. X} \textbf{13}, 041040 (2023).

\bibitem{Partanen2025a}
M.~Partanen and J.~Tulkki, \enquote{Gravity generated by four one-dimensional
  unitary gauge symmetries and the {S}tandard {M}odel,} \emph{Rep. Prog. Phys.}
  \textbf{88}, 057802 (2025).

\bibitem{Schwartz2014}
M.~D. Schwartz, \emph{Quantum Field Theory and the Standard Model}, Cambridge
  University Press, Cambridge (2014).

\bibitem{Peskin2018}
M.~E. Peskin and D.~V. Schroeder, \emph{An Introduction to Quantum Field
  Theory}, CRC Press, Boca Raton, FL (2018).

\bibitem{Partanen2025c}
M.~Partanen and J.~Tulkki, \enquote{Light deflection in unified gravity and
  measurable deviation from general relativity,}
  \emph{\normalfont{arXiv:2505.14446}}  (2025).

\bibitem{Feynman1961}
R.~P. Feynman, \emph{Quantum Electrodynamics}, W. A. Benjamin, New York (1961).

\bibitem{Landau1982}
V.~B. Berestetskii, E.~M. Lifshitz, and L.~P. Pitaevskii, \emph{Quantum
  Electrodynamics}, Pergamon, Oxford (1982).

\bibitem{Foldy1950}
L.~L. Foldy and S.~A. Wouthuysen, \enquote{On the dirac theory of spin 1/2
  particles and its non-relativistic limit,} \emph{Phys. Rev.} \textbf{78}, 29
  (1950).

\bibitem{Foldy1952}
L.~L. Foldy, \enquote{The electromagnetic properties of {D}irac particles,}
  \emph{Phys. Rev.} \textbf{87}, 688 (1952).

\bibitem{Shankar1994}
R.~Shankar, \emph{Principles of Quantum Mechanics}, Plenum Press, New York
  (1994).

\bibitem{Sakurai1994}
J.~J. Sakurai, \emph{Modern Quantum Mechanics}, Addison-Wesley, Reading, MA
  (1994).

\bibitem{Turyshev2007}
S.~G. Turyshev, M.~Shao, and K.~Nordtvedt, \enquote{Mission design for the
  laser astrometric test of relativity,} \emph{Adv. Space Res.} \textbf{39},
  297 (2007).

\bibitem{Turyshev2009b}
S.~G. Turyshev, M.~Shao, K.~L. Nordtvedt, H.~Dittus, C.~Laemmerzahl, S.~Theil,
  C.~Salomon, S.~Reynaud, T.~Damour, U.~Johann, P.~Bouyer, P.~Touboul,
  B.~Foulon, O.~Bertolami, and J.~Páramos, \enquote{Advancing fundamental
  physics with the laser astrometric test of relativity,} \emph{Exp. Astron.}
  \textbf{27}, 27 (2009).

\end{thebibliography}
\end{document}